\documentclass[lettersize,journal]{IEEEtran}
\usepackage{float}
\usepackage{amsmath,amsfonts}
\usepackage{algorithmic}
\usepackage{algorithm}
\usepackage{array}
\usepackage[caption=false,font=normalsize,labelfont=sf,textfont=sf]{subfig}
\usepackage{textcomp}
\usepackage{stfloats}
\usepackage{url}
\usepackage{xurl}
\usepackage{verbatim}
\usepackage{subcaption}
\usepackage{graphicx}
\usepackage{caption}
\usepackage{xcolor}
\usepackage{makecell}
\usepackage{multirow}
\usepackage{tablefootnote}
\usepackage{multicol}
\usepackage{latexsym}
\usepackage{booktabs}
\usepackage{tabularx} 
\usepackage{cite}
\usepackage[
    colorlinks=true,
    citecolor=blue,
    linkcolor=black,
    urlcolor=blue
]{hyperref}
\usepackage{orcidlink}
\newcolumntype{Y}{>{\raggedright\arraybackslash}X}

\begin{document}

\title{Human-Machine Bidirectional Trust-Aware Analysis and Design for Human-Led Truck Platooning}

\author{Chenzhao Li\textsuperscript{1}, Yunzhijun Yu\textsuperscript{2}, Yukun Lu\textsuperscript{3}\,\orcidlink{0000-0003-3259-5945}

\textsuperscript{1} Faculty of Interdisciplinary Studies, University of New Brunswick, Fredericton, NB E3B 5A3, Canada

\textsuperscript{2} Faculty of Management, University of New Brunswick, Fredericton, NB E3B 5A3, Canada

\textsuperscript{3} Dept. of Mechanical Engineering, University of New Brunswick, Fredericton, NB E3B 5A3, Canada
}


\maketitle

\begin{abstract}

Human-led truck platooning, where a human-driven truck leads one or more autonomous followers, offers significant benefits in fuel efficiency, safety, and traffic flow. However, its successful deployment hinges on trust between the human driver and the automated systems. Unlike conventional automation, trust in this context is inherently bidirectional: the human must trust the autonomous followers, and the followers must reliably interpret and respond to the human’s behavior. While prior research has extensively studied human trust in automation, the reciprocal nature of trust, especially considering the expertise of professional truck drivers, remains underexplored. This paper develops a conceptual framework of bidirectional trust for human-led platooning systems. Drawing on established trust theories (ability, benevolence, integrity) and insights from truck driver psychology, we propose distinct dimensions for human-to-automation trust and automation-to-human trust. To move beyond conceptualization, we introduce a quantitative model that operationalizes the bidirectional dynamics, using following distance as the key interaction variable to illustrate how trust co-evolves through a feedback loop. Simulation examples demonstrate both positive reinforcement and negative spiral effects. Based on this framework and its quantitative instantiation, we derive design guidelines for autonomous followers to foster appropriate trust calibration, improve safety, and enhance user acceptance. The framework bridges human factors and engineering perspectives, providing a theoretical and preliminary quantitative foundation for future empirical and modeling research.

\end{abstract}

\section{Introduction}

Human-led truck platooning, where a human-driven truck leads one or more autonomous followers, offers a promising approach to realizing the benefits of automation while maintaining human oversight in the freight sector \cite{alqahtani2025recent,lu2026rationale}. However, its successful deployment often depends critically on trust: not only the human driver's trust in the autonomous followers, but also the followers' capacity to reliably assess and respond to the human’s actions \cite{wendel2020trust}. In other words, trust in human-led platooning is inherently bidirectional.

While prior research has extensively examined human trust in automation, much of this literature treats trust as unidirectional (i.e., from human to machine) \cite{lee2004trust, hoff2015trust}. The reciprocal direction of trust,  especially how automated systems evaluate and calibrate their reliance on human operators, has received comparatively little attention \cite{garlichs2019trip}. This gap is particularly important for platooning, where the autonomous followers must continuously decide how closely to follow, when to intervene, and how much to rely on the human leader's decisions. 

We argue that human-to-automation (H2A) trust and automation-to-human (A2H) trust play structurally coupled but conceptually asymmetric roles. H2A trust is a psychological attitude shaped by perceived ability, benevolence, and integrity~\cite{lee2004trust,mayer1995integrative}. By contrast, A2H trust is not trust in the human psychological sense; rather, it denotes an automation-estimated human reliability or reliance score, computed from observable behavioral indicators such as driving consistency, rule compliance, and context-sensitive adaptation. We use the term A2H trust as a trust-like engineering construct because it governs how much the autonomous follower relies on the human leader's behavior when selecting following distance, intervention thresholds, or prediction confidence. Despite this asymmetry, the two processes are dynamically coupled: each party's behavior influences the other's evaluation, creating a feedback loop that can either stabilize or destabilize trust dynamics in the platoon.

Compared with non-professional drivers, professional truck drivers accumulate extensive expertise under demanding conditions and operate within distinctive safety, workload, and organizational constraints~\cite{duke2010age,castritius2021truck,lourencco2024representations}. At the same time, broader research on automated driving indicates that drivers differ in driving styles and expectations for automation behavior, and may prefer automated systems whose behavior is adapted to user preferences or perceived safety requirements~\cite{basu2017you,schrum2024maveric,nordhoff2023driver}. These findings suggest that an autonomous follower should not rely on a one-size-fits-all model of the human leader. It must not only demonstrate competence to earn the driver's trust~\cite{mayer1995integrative}, but also assess whether the leader's observed behavior reflects reliable professional adaptation or potentially unsafe behavior. Trust- and misbehavior-based platooning models provide a technical basis for such assessment, but they do not yet fully capture professional-driver expertise and context-dependent adaptation~\cite{garlichs2019trip}. Existing unidirectional trust models therefore remain insufficient for this setting, because they primarily explain how humans trust automation rather than how automation should evaluate and calibrate reliance on human leaders.


This paper makes three important contributions. First, we address this gap by developing a conceptual framework of bidirectional trust for human-led platooning, defining distinct dimensions for H2A trust (ability, benevolence, integrity)~\cite{mayer1995integrative} and A2H trust (consistency, rule compliance, context adaptability). Second, by incorporating psychological and behavioral characteristics of professional drivers \cite{duke2010age}, we introduce a quantitative model that operationalizes the bidirectional dynamics, using the following distance as a key interaction variable to illustrate how trust dynamically develops and evolves through a feedback loop. Third, based on the framework and the quantitative simulation, we derive design guidelines for autonomous followers to foster appropriate trust calibration and enhance safety. 

\section{Related Work}

\subsection{Trust in Automated Driving}

Trust has been defined as “an attitude that an agent will help achieve an individual’s goals under uncertainty”~\cite{lee2004trust}. A widely accepted model decomposes trustworthiness into three dimensions: ability (competence), benevolence (goodwill), and integrity (adherence to norms) \cite{mayer1995integrative}. In automated driving, these dimensions manifest through observable behaviors: smooth control demonstrates ability; cooperative spacing signals benevolence; proper signaling reflects integrity \cite{alqahtani2025recent,shugame}. Trust calibration, aligning a driver’s trust with system capabilities, is critical to avoid misuse (over-trust) or disuse (under-trust) \cite{alqahtani2025recent, hoff2015trust}. While extensive research has focused on how humans develop trust in automation, this literature predominantly treats trust as a one-directional phenomenon from human to machine \cite{lee2004trust, hoff2015trust}. In conventional automation, the machine's assessment of the human is often implicit or unnecessary, as the human is assumed to be the sole decision-maker. However, in human-led platooning, trust is inherently reciprocal: the autonomous follower is also required to evaluate the human leader.

\subsection{Truck Drivers: Expertise and Trust Expectations}

Professional truck drivers differ from ordinary drivers in experience, risk perception, and operational constraints~\cite{duke2010age}. They accumulate extensive expertise under challenging conditions, making them less tolerant of automation that deviates from established practices \cite{nordhoff2023driver}. Systems that mirror skilled driver strategies, such as gradual braking downhill, inspire greater confidence \cite{waytz2014mind}. Drivers also exhibit individual driving styles \cite{schrum2024maveric} and prefer automation that adapts a less aggressive version of their own style \cite{basu2017you}. Moreover, their adaptive strategies can sometimes border on risky behaviors, requiring the automated follower to distinguish between expert judgment and unsafe habits \cite{endsley2017here}. These characteristics imply that any trustworthy autonomous follower should not only demonstrate ability, benevolence, and integrity, but also be sensitive to the leader’s expertise and variability, a requirement that points toward bidirectional trust.

\subsection{Assessment of Human Reliability}

A small but growing body of research addresses the need for automation to assess human reliability. Game-theoretic models capture reciprocity and reputation in repeated human-machine interactions \cite{razin2021trust}, providing a foundation for bidirectional trust dynamics. In platooning, Garlichs et al. \cite{garlichs2019trip} proposed a “trust score” that enables followers to adjust following distance based on the predecessor’s consistency, a practical instantiation of automation evaluating human behavior. However, such models typically lack integration with psychological trust dimensions (ability, benevolence, integrity) and do not account for the specific expertise of truck drivers. This gap motivates our framework, which synthesizes insights from trust theory, driver psychology, and control perspectives to define both human-to-automation and automation-to-human trust dimensions tailored to human-led platooning.

\section{Bidirectional Trust Framework}

\begin{figure*}
    \centering
    \includegraphics[width=1\linewidth]{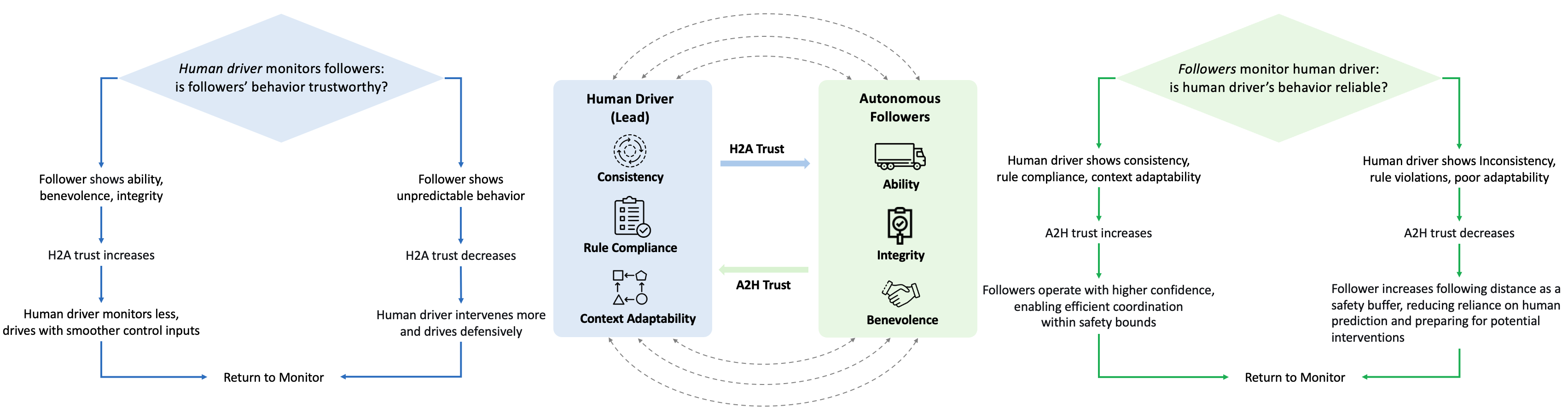}
    \caption{Bidirectional Trust Framework}
    \label{overview}
\end{figure*}

In human-led platooning, trust operates in two directions simultaneously. We distinguish two distinct constructs: human-to-automation trust (the human leader's trust in the autonomous follower) and automation-to-human trust (the follower's assessment of the human leader's reliability). For brevity, we refer to these as H2A trust and A2H trust, respectively, throughout the paper. Importantly, these constructs are conceptually asymmetric: H2A trust refers to the human leader's psychological trust in the autonomous follower, whereas A2H trust refers to the follower's operational estimate of the human leader's reliability. Thus, A2H trust should be understood as a trust-like reliance score rather than a subjective mental state. These two trust processes are not independent; they interact through a feedback loop, as illustrated in Fig. \ref{overview}. When the follower demonstrates trustworthy behavior, H2A trust increases, leading the human to reduce monitoring and drive more naturally, which in turn reinforces A2H trust. Conversely, erratic human behavior reduces A2H trust, prompting the follower to adopt a more cautious mode, which may affect H2A trust if the human perceives the follower as overly conservative. Drawing on established trust theories \cite{lee2004trust, mayer1995integrative} and insights from truck driver psychology \cite{duke2010age}, we propose distinct dimensions for each direction of trust, tailored to the platooning context.

\subsection{Human-to-Automation Trust (H2A)}

For the human leader to trust an autonomous follower, the follower should demonstrate three qualities adapted from the ability-benevolence-integrity framework \cite{mayer1995integrative}. Ability refers to the follower's competence in executing driving tasks, manifesting as smooth acceleration and braking, precise lane-keeping, and appropriate responses to traffic conditions; professional drivers, with their extensive experience, are particularly sensitive to deviations from competent behavior~\cite{nordhoff2023driver}, and a follower that brakes too early or too late may quickly lose trust \cite{waytz2014driving}. Benevolence captures the perception that the follower acts cooperatively and with goodwill, evidenced by maintaining safe following distances, yielding to merging vehicles, and avoiding maneuvers that endanger others \cite{alqahtani2025recent}. Integrity concerns adherence to rules and norms following traffic laws, using signals appropriately, and behaving predictably which in platooning also implies transparency through clear communication of intentions \cite{mayer1995integrative}. These three dimensions collectively shape H2A trust, though their relative importance may vary across drivers and situations \cite{basu2017you}.

\subsection{Automation-to-Human Trust (A2H)}
\label{A2H}

For the autonomous follower to reliably depend on the human leader's actions for safe coordination, it should evaluate the leader's behavior along three dimensions analogous to, but distinct from, those used by humans. Consistency refers to the predictability of the human's driving behavior, stable speeds, smooth steering inputs, and predictable patterns in lane changes and braking, enabling the follower to anticipate future actions and plan accordingly \cite{garlichs2019trip}. Rule compliance captures the extent to which the human adheres to traffic laws and driving norms, including proper use of turn signals, observance of speed limits, and respect for right-of-way. Context adaptability reflects the human's ability to adjust driving behavior appropriately to situational demands, such as increasing following distance in rain or modulating speed on downhill grades; the follower can recognize such adaptations as markers of expertise rather than erratic behavior, while remaining alert to actions that deviate from safe operational norms \cite{endsley2017here}. While these A2H dimensions are not intended as direct analogs of the ABI framework, they reflect the specific behavioral indicators that an autonomous follower requires to dynamically assess the leader's reliability for safe coordination, and enables the follower to modify its own behavior accordingly.

\subsection{Bidirectional Dynamics}

H2A and A2H trust co-evolve through ongoing interaction. When the follower demonstrates high ability, benevolence, and integrity, the human's trust increases. This leads the human to monitor less and drive more naturally. In turn, if the human's driving remains consistent and rule-compliant, the follower's trust in the human also increases. This positive feedback loop stabilizes the platoon in a state of high mutual trust. On the other hand, a negative loop may arise: if the follower behaves unpredictably, H2A trust declines, prompting the human to intervene more frequently or drive defensively; this may manifest as erratic inputs that reduce A2H trust, causing the follower to adopt even more conservative behavior, further eroding the human's confidence. Game-theoretic models of repeated interaction capture such dynamics through concepts of reputation and reciprocity~\cite{axelrod1981evolution, trivers1971evolution}, providing a formal foundation for the bidirectional trust processes described here.

\section{Design Guidelines for Trustworthy Autonomous Followers}

\begin{table*}[t]
\caption{Implementation matrix mapping trust dimensions to design objectives, implementation approaches, enabling technologies, and evaluation metrics.}
\label{tab:trust-implementation-matrix}
\newcolumntype{Y}{>{\raggedright\arraybackslash}X}
\renewcommand{\arraystretch}{1.2}
\begin{tabularx}{\textwidth}{l l l Y Y}
\toprule
\textbf{Trust Dimension} & \textbf{Design Objective} & \textbf{Implementation Approach} & \textbf{Enabling Technology} & \textbf{Evaluation Metric} \\
\midrule
\textbf{H2A Ability} &
Smooth driving &
Jerk-minimizing control &
MPC, RL &
\makecell[l]{RMS jerk; \\ Lane deviation} \\ [8pt]

\textbf{H2A Benevolence} &
Cooperative behavior &
Dynamic headway adjustment &
Adaptive control &
\makecell[l]{Yielding frequency; \\ Time-to-collision} \\ [8pt]

\textbf{H2A Integrity} &
Transparency &
Intent display on HUD &
\makecell[l]{AR-HUD, \\ V2V communication} &
\makecell[l]{Driver comprehension time; \\ Mental load} \\ [8pt]

\textbf{A2H Consistency} &
Predictability assessment &
Real-time consistency scoring &
\makecell[l]{Time-series analysis, \\ Bayesian inference} &
\makecell[l]{Prediction error; \\ Trust score stability} \\ [8pt]

\textbf{A2H Rule Compliance} &
Violation detection &
Visual perception &
\makecell[l]{Multimodal fusion, \\ Computer vision} &
\makecell[l]{Detection latency; \\ False positive rate} \\ [8pt]

\textbf{A2H Context Adaptability} &
Expert vs.\ risk distinction &
Expert driver behavior modeling &
\makecell[l]{Supervised learning, \\ Anomaly detection} &
\makecell[l]{Classification accuracy; \\ False alarm rate} \\
\bottomrule
\end{tabularx}
\end{table*}

Based on the bidirectional trust framework proposed above, we derive design guidelines for autonomous followers in human-led platooning. These guidelines translate each trust dimension into concrete design requirements, organized around the two directions of trust.

\subsection{Designing for Human-to-Automation Trust}

To earn and maintain the human leader's trust, the autonomous follower should demonstrate ability, benevolence, and integrity through its driving behavior and communication.

\textbf{Demonstrating ability} requires the follower to execute driving tasks with smoothness and precision. Control algorithms should minimize jerk during acceleration and braking, maintain stable lane positioning, and respond appropriately to traffic events such as cut-ins or sudden decelerations. Drawing on insights from professional drivers \cite{waytz2014driving}, the follower should adopt braking and acceleration profiles that mirror those of experienced truck drivers, for example, gradual braking on downhill grades rather than abrupt deceleration. Implementing adaptive cruise control with naturalistic acceleration patterns signals competence and reduces the likelihood of the human perceiving the follower as inept \cite{craig2021should}.


\textit{Signaling benevolence} involves cooperative behaviors that demonstrate goodwill toward the human leader and other road users. The follower should maintain following distances that are safe while remaining compatible with professional drivers' gap preferences and perceived comfort in platooning contexts \cite{castritius2021truck,nodine2017naturalistic}. When other vehicles attempt to merge, the follower should yield appropriately rather than contesting the space, thereby prioritizing safety and traffic cooperation over aggressive efficiency. These cooperative actions can help the human leader interpret the follower as predictable and safety-oriented. External human-machine interfaces (e.g., light-based intent displays or external displays) may further communicate the follower's state and intended behavior to surrounding road users, particularly in ambiguous interaction situations \cite{rouchitsas2019external}.

\textbf{Ensuring integrity and transparency} requires consistent rule adherence and clear communication of intentions. The follower should use turn signals when changing lanes or merging, respect speed limits, and follow right-of-way rules. Beyond rule compliance, the follower should make its planned actions and current operating state known to the human leader through intuitive in-cab interfaces, for example, a head-up or dashboard display indicating ``preparing to slow down'' or ``maintaining distance.'' Prior work on automated-vehicle transparency shows that communicating system status, intended behavior, and environmental hazards can support trust calibration, usability, and situation awareness~\cite{kraus2020more,oliveira2020influence}. Predictable behavior, such as consistently signaling before a lane change, further reinforces perceived integrity by enabling the human driver to anticipate the follower's actions, thereby reducing uncertainty and supporting appropriate reliance~\cite{lee2004trust,mayer1995integrative}.

\subsection{Designing for Automation-to-Human Trust}

For the follower to develop accurate assessments of the human leader's reliability and adjust its behavior accordingly, it should continuously evaluate consistency, rule compliance, and context adaptability.

\textbf{Assessing consistency} involves monitoring the predictability of the human's driving behavior. The follower can track metrics such as speed variability, steering smoothness, and lane position stability over time \cite{garlichs2019trip}. When the leader's behavior is highly consistent, the follower can operate with tighter following distances and more confident predictions. If inconsistency is detected, frequent speed fluctuations or abrupt steering corrections, the follower should increase following distance and prepare for potential interventions. Implementing a real-time ``consistency score" enables dynamic adaptation to the leader's demonstrated reliability.

\textbf{Detecting rule compliance} requires the follower to monitor whether the human adheres to traffic laws and driving norms. This includes detecting proper use of turn signals, observance of speed limits, and compliance with lane discipline~\cite{madhavan2007effects}. Sensors and perception algorithms can recognize these behaviors; for instance, camera systems can track whether the leader signals before lane changes. When repeated violations are detected, such as frequent failures to signal, the following vehicle should adjust its trust estimate downward and transition to a more conservative driving mode, thereby increasing safety margins.

\textbf{Recognizing context adaptability} is the most nuanced dimension, requiring the follower to distinguish adaptive expertise from genuinely risky behavior. The following vehicle should integrate situational context, including weather conditions, road geometry, and traffic density, into its evaluation of the human driver’s actions \cite{endsley2017here}. Increased following distance in rain may indicate prudent adaptation, while excessive speed in fog signals poor judgment. Machine learning models trained on expert driver data could enable the follower to recognize context-appropriate adaptations as markers of expertise, while flagging deviations from safe norms as trust-reducing events. When context-inappropriate behavior is detected, the follower should increase following distance and be prepared to intervene if necessary.

\section{Illustrative Scenarios}

This section presents two typical platooning scenarios to illustrate how the bidirectional trust framework and design guidelines apply in practice. These examples demonstrate the interplay between H2A and A2H trust in real-world driving situations.

\subsection{Highway Following}

Consider a human-led platoon traveling on a highway at steady speed. The autonomous follower maintains a constant following distance, responds smoothly to the leader's speed adjustments, and signals clearly before any lane changes. Over time, the human leader observes that the follower handles routine driving tasks competently, accelerating and braking smoothly, holding its lane precisely, and reacting appropriately to surrounding traffic. This consistent demonstration of ability gradually increases the human's H2A trust. The follower also demonstrates integrity through transparent communication: a head-up display shows the follower's intentions: ``maintaining distance" or ``preparing to slow", allowing the human to anticipate its actions and reinforcing the perception that the follower operates reliably in accordance with expected norms.

As trust grows, the human reduces monitoring and drives more naturally, resulting in smoother and more predictable driving. The follower's sensors detect this increased consistency, stable speed, smooth steering inputs, predictable lane positioning which raises its A2H trust in the human. With higher A2H trust, the follower can safely maintain a tighter following distance, improving fuel efficiency and traffic flow. This positive feedback loop stabilizes the platoon in a high-trust equilibrium, benefiting both parties through reduced cognitive load and improved operational efficiency. It is worth noting that individual driver preferences vary: a driver with a more aggressive style might prefer closer following distances, while a safety-conscious driver may value larger gaps. A trustworthy follower should adapt within safe bounds to accommodate such preferences.

\subsection{Ramp Merging}

A more complex scenario involves the platoon approaching a highway on-ramp where a merging vehicle seeks to enter. If the human leader chooses to yield by slowing slightly to allow a merging vehicle to enter, the following vehicle should interpret this behavior as a cooperative action and respond accordingly by maintaining space and yielding as well. This cooperative behavior signals benevolence to the human, reinforcing H2A trust, as the human observes that the follower shares their cooperative driving values and responds predictably. At the same time, the following vehicle assesses the human’s merging decision. A smooth and well-timed yield indicates strong contextual awareness, which increases the follower’s autonomy-to-human trust and reinforces the perception that the human can be relied upon to make sound judgments in complex traffic situations.

However, if the human leader attempts to merge at an unsafe moment or fails to signal, the follower's sensors detect this rule compliance issue and lower its A2H trust estimate accordingly, reflecting increased uncertainty about future actions. In response, the follower increases following distance and prepares for potentially erratic behavior. This conservative mode may be perceived by the human as overly cautious, potentially reducing H2A trust if the reasoning is not understood. Here, transparency becomes essential: if the following vehicle communicates its reasoning, for example by displaying a dashboard message indicating that it has detected uncertain driver behavior and is increasing distance, the human driver may better understand the system’s response and maintain trust despite its more conservative behavior. This illustrates how integrity and transparency support trust even in challenging situations.

\subsection{Summary}

These scenarios illustrate how the bidirectional trust framework operates in practice. The highway following example shows positive trust dynamics reinforcing efficient platoon operation, while highlighting the role of transparency and individual driver preferences. The ramp-merging example illustrates how situational factors and human decisions shape trust in both directions, and how the follower’s adaptive responses, together with clear communication, can either preserve or undermine trust depending on the context. Together, they underscore the importance of designing autonomous followers that not only demonstrate trustworthiness across all dimensions but also accurately assess human reliability and adjust their behavior accordingly.

\section{Quantitative Illustration of Bidirectional Trust Dynamics with Distance-Mediated Interaction}
\label{sec:quant}

To concretize the bidirectional trust framework and illustrate its dynamic coupling, we develop a minimal dynamical model where the following distance serves as the key interaction variable. Let \(T_{\text{H2A}}(t) \in [0,1]\) and \(T_{\text{A2H}}(t) \in [0,1]\) denote the human's trust in the autonomous follower and the follower's trust in the human, respectively, at discrete time~\(t\).

\subsubsection{Trust-Based Gap Adjustment}
The autonomous follower adjusts its desired following distance based on its current trust in the human driver:
\begin{equation}
d(t) = d_{\text{min}} \left[ 1 + \kappa \bigl(1 - T_{\text{A2H}}(t)\bigr) \right]
\label{eq:gap}
\end{equation}
where \(d_{\text{min}}\) is the \emph{minimum safe following distance} maintained when trust is highest (e.g., based on a 1-second time headway as introduced in SAE J2944), and \(\kappa > 0\) scales the maximum additional gap when trust is lowest. This formulation captures the intuition that higher trust allows tighter following, while lower trust demands larger safety margins.

\subsubsection{Human Perception of Distance}
The human driver observes the actual gap \(d(t)\) and compares it with an expected distance \(d_{\text{exp}}\), which may reflect individual driving style preferences (e.g., a more cautious driver may expect a larger gap). The deviation from expectation influences the human's assessment of the follower's trustworthiness. We define the human's observation score for H2A trust as:
\begin{equation}
O_{\text{H2A}}(t) = 1 - \gamma \left( \frac{d(t) - d_{\text{exp}}}{d_{\text{exp}}} \right)^2,
\quad \text{clipped to }[0,1]
\label{eq:obsH2A}
\end{equation}
with sensitivity parameter \(\gamma > 0\). The quadratic term penalizes both too-close and too-far gaps symmetrically.

\subsubsection{Follower's Observation of Human Driving}
The driving quality score \(q(t)\) reflects the follower's real-time assessment of the human's driving behavior along the A2H trust dimensions defined in Section~\ref{A2H} (consistency, rule compliance, and context adaptability). Higher values indicate smoother steering, more stable speed, proper use of turn signals, and appropriate adaptation to traffic conditions, all markers of predictable and trustworthy driving. Conversely, lower values correspond to erratic maneuvers, frequent speed fluctuations, missing signals, or context-inappropriate actions. The follower continuously monitors the smoothness and predictability of the human's driving. For simplicity in this illustrative model, we assume that driving quality improves when the human is more trusting and relaxed. Let \(q(t) \in [q_{\text{min}}, q_{\text{max}}]\) represent the observed driving quality at time \(t\), modeled as a linear function of \(T_{\text{H2A}}(t)\):
\begin{equation}
q(t) = q_{\text{min}} + (q_{\text{max}} - q_{\text{min}})\, T_{\text{H2A}}(t)
\label{eq:quality}
\end{equation}

The follower's observation score for A2H trust is then set to the normalized quality:
\begin{equation}
O_{\text{A2H}}(t) = q(t)
\label{eq:obsA2H}
\end{equation}

Note that this positive correlation may not hold for all drivers; some individuals might exhibit degraded performance when overly relaxed due to reduced vigilance. Capturing such heterogeneity would require modeling individual driver traits, which we leave for future work.

\subsubsection{Trust Update Dynamics}
Both trust values evolve according to exponential smoothing (leaky integrator) to capture the persistence of trust over time:
\begin{align}
T_{\text{H2A}}(t+1) &= \alpha \, T_{\text{H2A}}(t) + (1-\alpha) \, O_{\text{H2A}}(t) \label{eq:updateH2A} \\
T_{\text{A2H}}(t+1) &= \beta \, T_{\text{A2H}}(t) + (1-\beta) \, O_{\text{A2H}}(t) \label{eq:updateA2H}
\end{align}
where \(\alpha, \beta \in (0,1)\) control the update rates (higher values mean slower adaptation).

\begin{table}[ht]
\caption{Model parameters and initial trust settings.}
\begin{tabular}{lll}
\toprule
\textbf{Parameter} & \textbf{Description} & \textbf{Value}\\
\midrule
$v$ & Vehicle speed (km/h) & Variable \\
$d_{\text{min}}$ &  $v$/3.6 (m) & Variable \\
$d_{\text{exp}}$ & $v * t_{\text{exp}}$ (m) & Variable \\
$\kappa$ & Scale gain & $0.5$ \\
$\gamma$ & Sensitivity parameter & $2$ \\
$q_{\text{min}}$ & Min. drive quality score & $0.5$ \\
$q_{\text{max}}$ & Max. drive quality score & $1$ \\
$\alpha$ & H2A update rate gain  & $0.8$ \\
$\beta$ & A2H update rate gain & $0.8$ \\
$T_{\text{H2A}}(0)$ & Initial H2A trust level & $0.8$ (positive) \& $0.6$ (negative) \\
$T_{\text{A2H}}(0)$ & Initial A2H trust level & $0.8$ (positive) \& $0.6$ (negative) \\
\bottomrule
\end{tabular}
\label{table:sim_paras}
\end{table}

\section{Preliminary Simulation Studies}

\begin{figure}
    \centering
    \includegraphics[width=0.9\linewidth]{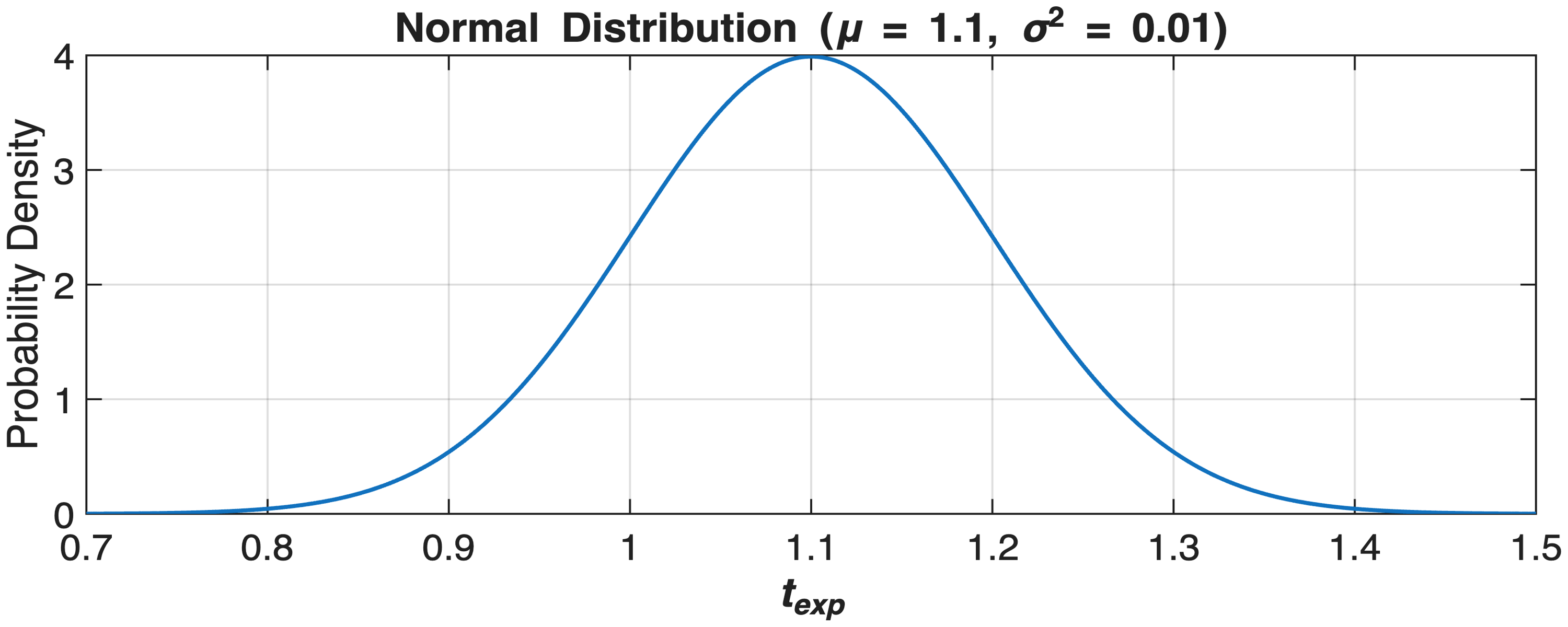}
    \caption{Human driver's expected time headway $t_{\text{exp}}$.}
    \label{fig:t_exp}
\end{figure}

We simulate two scenarios, positive and negative, to illustrate the bidirectional trust feedback mechanism. 
The vehicle speed $v$ is modeled as a uniformly distributed random variable within the range $[98, 102]$ km/h. 
The minimum safe following distance $d_{\text{min}}$ follows the one-second time-headway rule. 
The human driver's expected following distance $d_{\text{exp}}$ is computed based on the vehicle speed and the driver's expected time headway $t_{\text{exp}}$. 
To capture uncertainty and variability in human judgment, the expected time headway $t_{\text{exp}}$ is modeled as a Gaussian random variable  $\mathcal{N}(\mu = 1.1,\ \sigma^2 = 0.01)$, as illustrated in Fig.~\ref{fig:t_exp}. The parameters used in this simulation study are described in Table~\ref{table:sim_paras}.

\subsection{Positive Scenario}
\begin{figure}
    \centering
    \includegraphics[width=\linewidth]{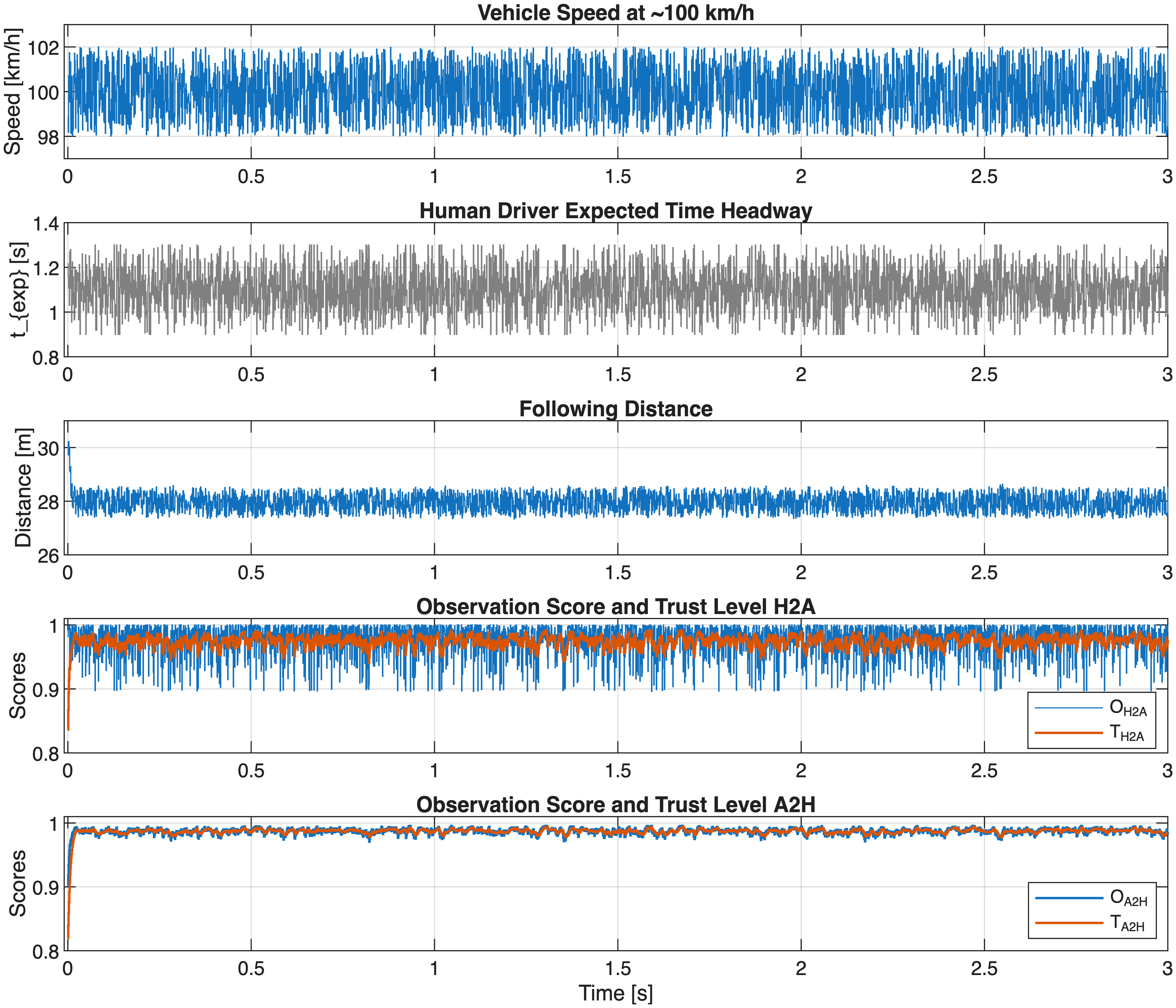}
    \caption{Positive scenario simulation.}
    \label{fig:positive}
\end{figure}

Fig.~\ref{fig:positive} illustrates the simulation results for the positive scenario over a 3-second horizon. The vehicle speed fluctuates around 100 km/h, remaining within the predefined uniform range. The human driver’s expected time headway varies around its Gaussian mean, reflecting stochastic variability in human judgment. As a result, the computed following distance stabilizes around 28 m, with minor fluctuations driven by speed and time-headway variations.

The fourth subplot presents the observation score $O_{H2A}$ and the corresponding trust level $T_{H2A}$. Both remain consistently high (close to 1), indicating that the automation behaves smoothly and maintains appropriate following behavior relative to the human driver’s expectations. Although the observation score exhibits small stochastic variations, the trust level remains stable, demonstrating gradual trust dynamics.

Similarly, the fifth subplot shows the observation score $O_{A2H}$ and trust level $T_{A2H}$. Both metrics rapidly converge to and remain near unity, suggesting that the automation perceives the human driver as consistent and reliable under this positive scenario. Overall, the bidirectional trust dynamics remain stable and high, illustrating mutually reinforced confidence between the human and the automation in nominal operating conditions.

\subsection{Negative Scenario}
A human error is applied in this scenario, modeled by temporarily dropping \(q(t)\) to \(0.3\) at \(t=1\) that reduces \(O_{\text{A2H}}\). This lowers \(T_{\text{A2H}}\) and widens the gap via Eq.~\eqref{eq:gap}. The human perceives the enlarged gap as overly cautious (low \(O_{\text{H2A}}\)), reducing \(T_{\text{H2A}}\) and further degrading driving quality. This negative loop amplifies the initial perturbation.

\begin{figure}
    \centering
    \includegraphics[width=\linewidth]{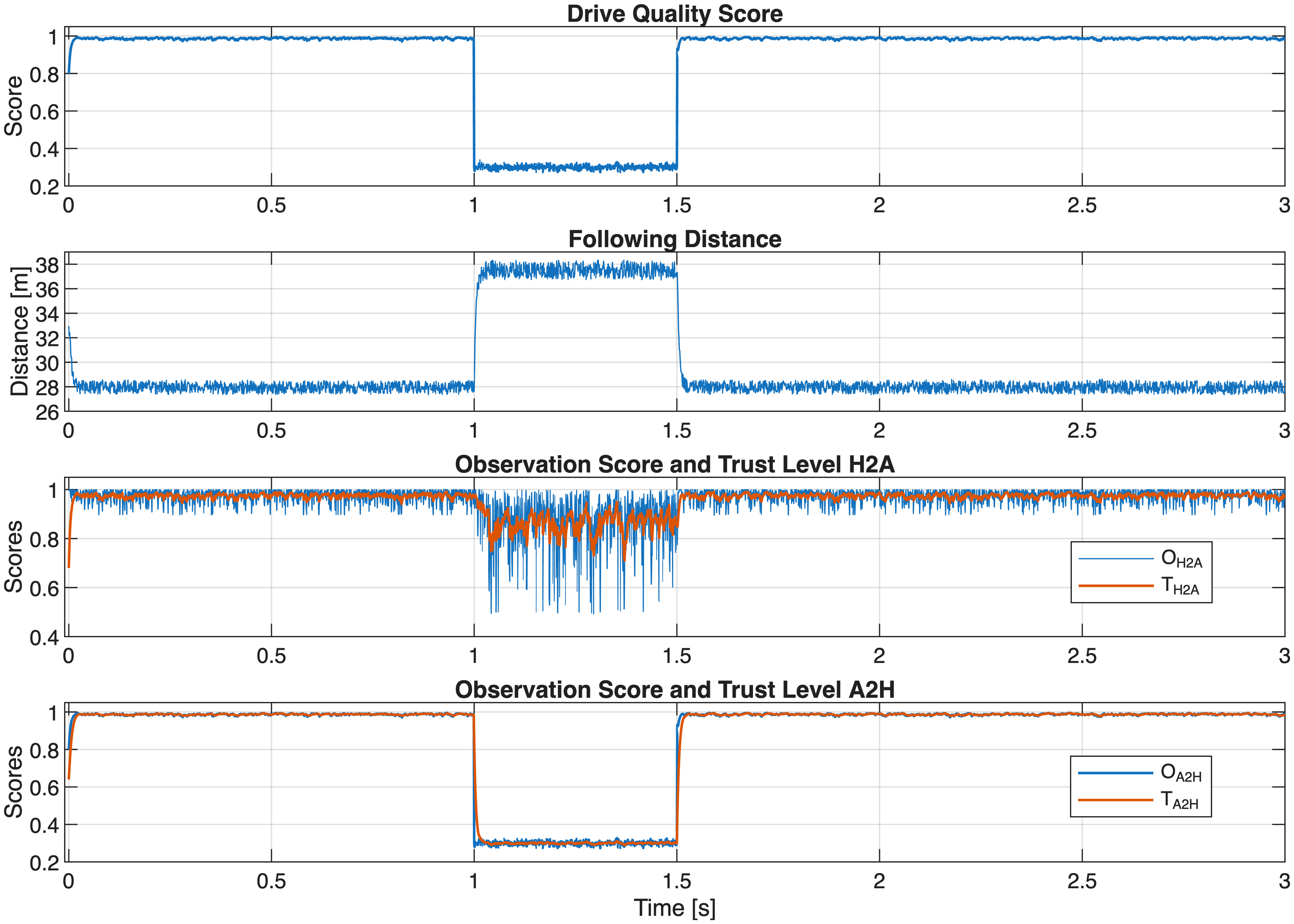}
    \caption{Negative scenario simulation.}
    \label{fig:negative}
\end{figure}

Fig.~\ref{fig:negative} presents the simulation results for the negative scenario, where a temporary mismatch between expected and actual following behavior is introduced between 1.0 s and 1.5~s. The drive quality score (first subplot) drops sharply from near 1 to approximately 0.3 during this interval, indicating degraded driving performance. Correspondingly, the following distance (second subplot) increases significantly from its nominal value of approximately 28 m to around 37-38 m, reflecting an abnormal or overly conservative headway relative to expectation. After 1.5~s, both the drive quality score and following distance return to nominal levels.

The third subplot shows the observation score $O_{H2A}$ and trust level $T_{H2A}$. During the degraded interval, $O_{H2A}$ exhibits substantial fluctuations and pronounced drops, while $T_{H2A}$ decreases moderately, demonstrating a filtered and gradual trust adaptation rather than instantaneous collapse. Once normal behavior resumes, trust progressively recovers toward its previous high level, illustrating the inertia and memory effect embedded in the trust dynamics.

Similarly, the fourth subplot presents $O_{A2H}$ and $T_{A2H}$. When the abnormal driving quality is detected, both metrics rapidly decrease to low values (approximately 0.3), indicating that the automation perceives the human driver as unreliable under these conditions. After the disturbance period ends, both the observation score and trust level quickly recover to near unity.

Overall, the negative case demonstrates the sensitivity of the bidirectional trust model to performance degradation. Temporary behavioral inconsistencies lead to measurable trust reductions in both directions, while the recovery phase highlights the system’s capability to restore trust once normal driving conditions are reestablished.

This quantitative illustration, though simplified, reveals several insights. First, the coupling via distance creates a feedback loop that can either reinforce or undermine trust. Second, the parameters $\alpha$ and $\beta$ govern the “memory” of trust: longer memory smooths transient errors but may delay recovery from persistent issues. Third, individual preferences (e.g., \(d_{\text{exp}}\)) modulate the human's sensitivity to the follower's behavior. The model provides a foundation for designing adaptive lead-follower controllers that not only respond to trust but also actively shape it through transparent communication and predictable behavior. Future work will embed this framework into model predictive control (MPC) to optimize both safety and trust efficiency.

\section{Discussion and Conclusion}

This paper has proposed a bidirectional trust framework for human-led truck platooning, arguing that trust in this context operates in two directions: the human leader must trust the autonomous follower, and the follower must continuously assess the human's reliability. Drawing on established trust theories \cite{lee2004trust} and insights from professional truck driver psychology \cite{duke2010age}, the framework defines distinct dimensions for human-to-automation trust (ability, benevolence, integrity) and automation-to-human trust (consistency, rule compliance, context adaptability). These dimensions capture the behavioral indicators that shape trust on both sides and highlight the dynamic interplay between them. The framework extends existing trust research by explicitly modeling reciprocity in human-machine interaction. It also bridges psychological constructs with engineering design: the proposed dimensions translate into observable behaviors, such as jerk for ability, headway for benevolence, and signal usage for integrity, that can be measured and incorporated into control algorithms \cite{garlichs2019trip}.

To move beyond a conceptual treatment, we introduced a quantitative illustration of the bidirectional dynamics mediated by following distance (Section~\ref{sec:quant}). This minimal model demonstrates how trust co-evolves through a feedback loop: the follower's trust determines the gap, which in turn influences the human's perception and behavior. Simulation examples reveal that the coupling can either reinforce a high-trust equilibrium or amplify perturbations into a downward spiral, underscoring the need for carefully calibrated interaction strategies.

Based on the framework, we derived design guidelines for autonomous followers, emphasizing the importance of transparent communication, adaptive behavior, and real-time assessment of human driving patterns. Transparency emerges as a critical enabler of trust calibration: when followers communicate their reasoning through intuitive interfaces, drivers are better able to understand and anticipate their actions, reducing uncertainty and preventing misinterpretation \cite{alqahtani2025recent}. The A2H dimensions suggest that followers should be equipped with monitoring capabilities to assess human consistency, rule compliance, and context adaptability, adjusting following distance and intervention thresholds accordingly \cite{garlichs2019trip}. Illustrative scenarios demonstrated how these dynamics play out in typical platooning situations, showing both positive feedback loops that reinforce efficient operation and challenges that can erode trust when behavior deviates from expectations.

The framework and its quantitative instantiation open several avenues for future research. First, empirical validation is essential: future driving simulator and on-road studies should examine whether the proposed dimensions capture professional drivers' trust, perceived safety, gap preferences, and behavioral adaptation in human-led platooning~\cite{hjalmdahl2017driver,castritius2021truck,botelho2025simulator}. Second, the dynamical model can be extended by replacing its simplifying assumptions (e.g., linear quality-trust relationship) with data-driven models learned from sensor data. Third, model parameters could be personalized to accommodate individual driving styles while maintaining safety. The current model assumes a positive correlation between human trust and driving quality; future work should relax this assumption by incorporating driver-specific traits (e.g., skill level, risk propensity) to capture scenarios where increased relaxation leads to degraded performance. Fourth, incorporating game-theoretic elements \cite{lin2025scenario} or or Bayesian networks would allow reasoning under uncertainty. Finally, the framework should be extended to include the trust of surrounding road users. By recognizing trust as a two-way street, this work provides a theoretical and preliminary quantitative foundation for designing platooning systems that are not only safe and efficient but also perceived as trustworthy partners by the humans who lead them.

\section{Future Work}
Future work may further enrich the dynamical assumptions underlying the quantitative illustration. The current negative scenario focuses on a short-term disturbance followed by recovery, serving as an initial demonstration of how bidirectional trust estimates respond to behavioral changes. Building on this formulation, future models could examine a broader range of interaction patterns, including longer-lasting disturbances, asymmetric trust recovery, and possible low-trust steady states. In addition, the present model adopts a simple positive relationship between H2A trust and observed driving quality for interpretability. Future extensions could incorporate driver heterogeneity, as trust may influence driving behavior differently depending on workload, vigilance, risk propensity, and professional experience. Such extensions would allow the model to represent richer trust dynamics while preserving the transparency of the current framework.

\bibliographystyle{IEEEtran}
\bibliography{bibfile}

\end{document}